\documentclass[a4paper,11pt]{article}
\usepackage{amsmath}
\usepackage{amsfonts}
\usepackage{mathbbol}
\usepackage{mathrsfs}
\numberwithin{equation}{section}
\usepackage{setspace}
\usepackage{hyperref}
\usepackage{xcolor}
\usepackage{cite}
\usepackage[top=1 in,bottom=1 in,left=1.0 in,right=1.0 in]{geometry}

\newtheorem{definition}{Definition}[section]

\title{Nonlinearization of the KdV-type and mKdV-type bilinear equations}
\author{Xin Zhang$^1$,~~ Jin Liu$^1$,~~ Da-jun Zhang$^{1,2}$\footnote{
Corresponding author. Email: djzhang@staff.shu.edu.cn}\\
{\small $^1$Department  of Mathematics, Shanghai University, Shanghai 200444, China}\\
{\small $^{2}$Newtouch Center for Mathematics of Shanghai University,  Shanghai 200444, China}}

\date{\today}

\begin{document}
	\noindent
	
\maketitle

\begin{abstract}
In this paper, we show a general procedure to nonlinearize bilinear equations by using the Bell polynomials.
As applications, we obtain nonlinear forms of some integrable bilinear equations (in the sense having 3-soliton solutions)
of the KdV type and mKdV type that were found by Jarmo Hietarinta in 1980s.
Examples of non-integrable bilinear equations of the KdV type are also given.

PACS Number(s): 02.30.Ik, 02.30.Ks, 05.45.Yv

\end{abstract}

\section{Introduction}\label{sec-1}

It is well known that Ryogo Hirota's bilinear method \cite{Hir-1971,Hirota-book}
is a powerful tool in the study of integrable systems.
With regard to finding solutions, nonlinear equations can be transformed into bilinear forms
and then soliton solutions can be derived.
According to the bilinear forms, bilinear equations were classified as the
Korteweg-de Vries (KdV) type, the modified KdV (mKdV) type, the sine-Gordon (sG) type
and the nonlinear Schr\"odinger (NLS) type by Jarmo Hietarinta \cite{Hie-1987a,Hie-1987b,Hie-1987c,Hie-1988}.
It is Hirota who first realized that the KdV-type bilinear equations
always have 1-soliton and 2-soliton solutions \cite{Hir-1980}
but having a 3-soliton solution indicates a kind of integrability,
which is now known as the integrability of bilinear equations in Hirota's sense \cite{Hir-1980,Hirota-book,Hie-1987a}.
In a series of celebrated papers in 1987 and 1988, Hietarinta examined the KdV-type,
mKdV-type and sG-type bilinear equations that have 3-soliton solutions \cite{Hie-1987a,Hie-1987b,Hie-1987c}
and the NLS-type bilinear equations that have 2-soliton solutions \cite{Hie-1988}.
Some new bilinear equations integrable in Hirota's sense were found
and so far bilinear forms of some of these equations are still not known.

It is also well known that bilinear equations have remarkable mathematical structures.
Sato found that the bilinear  Kadomtsev-Petviashvili (KP) hierarchy are the Pl\"ucker relations on the
infinitely dimensional Grassmannians \cite{Sato-1981},
which has led to a set of beautiful and profound theory for integrable systems
developed by the Kyoto group.
On the other hand, Lambert, Gilson and their collaborators found connections between
Hirota's bilinear derivatives and the Bell polynomials \cite{gilson-1996,lambert-1994},
which led to a mechanism to bilinearize nonlinear equations.
This mechanism was later extended to deriving bilinear B\"acklund transformations
and Lax pairs, etc \cite{Lambert-IP-2001,Lambert2001,LamS-2006}.
It was also extended to the study of supersymmetric systems \cite{fan-2012}
and can be implemented using symbolic computations \cite{miao-2014}.
Note that the Bell polynomials approach is in principle applicable for the KdV type equations
(see \cite{fan-2011,fan-2011-1,Luo-PLA-2011} as examples),
while the application to the mKdV-type, sG-type and NLS-type equations was less developed.

In this paper, we will describe a procedure to convert bilinear equations into their nonlinear forms.
This will enable us to obtain nonlinear forms of the new integrable bilinear equations found by Hietarinta
in \cite{Hie-1987a,Hie-1987b}.
Our procedure is also based on the Bell polynomials.
In fact, in \cite{gilson-1996,lambert-1994}, the connection between Hirota's bilinear derivatives and
the Bell polynomials has been revealed. However, in \cite{gilson-1996,lambert-1994}, all attention was paid
to bilinearize nonlinear equations rather than the reverse direction.
In this paper, we explain how the Bell polynomial approach works in nonlinearization bilinear equations.

The paper  is organized as follows.
First, in Sec.\ref{sec-2}, we recall the Bell polynomials, binary Bell polynomials and
their connections with Hirota's bilinear derivatives.
In Sec.\ref{sec-3}, we show how such a nonlinearization technique works
in the KdV-type and mKdV-type equations.
Illustrative examples include integrable equations and also non-integrable ones.
As a result, we obtain nonlinear forms of some integrable bilinear equations found by Hietarinta
in  \cite{Hie-1987a,Hie-1987b}.
Finally, concluding remarks are given in Sec.\ref{sec-4}.

\section{Bell polynomials theory}\label{sec-2}

In this section, we review the connection between the Bell polynomials and Hirota's bilinear derivatives.
One can also refer to the pioneer work \cite{gilson-1996,lambert-1994}.

\subsection{Bell polynomials}\label{sec-2-1}

The Bell polynomials of our interest are defined as the following.	

\begin{definition}\label{def-2-1}
 \cite{bell-1934} Let $h=h(x)$ be a $\mathrm{C}^\infty$ function of $x\in \mathbb{R}$ and
denote $h_r=\partial^{r}_{x} h$ for $r=1,2,\cdots$.
Then,  		
\begin{equation}\label{Y-def}
			Y_{[nx]}(h) \doteq Y_n(h_1, h_2, \cdots, h_n) = {\mathrm{e}}^{-h}\partial^{n}_{x}{\mathrm{e}}^{h}, ~~~~~
			n=1,2,\cdots
\end{equation}
are known as the Bell polynomials.
\end{definition}

In 1934, E.T. Bell \cite{bell-1934} studied these polynomials for several special  choices of $h(x)$.
These polynomials can be generated recursively by
\begin{equation}\label{Y-p-recur}
		Y_0=1, ~~ Y_{n+1}(h_1,h_2,\cdots,h_{n+1})=(\partial_x+h_1)Y_n(h_1,h_2,\cdots,h_n),
\end{equation}
which agrees with the (potential) Burgers hierarchy (see \cite{gilson-1996,lambert-1994}).
They can also be represented with a single formula
by means of the Arbogast formula \cite{Arbogast-1800}
or the Fa$\Grave{\text{a}}$ di Bruno formula \cite{Faa-1855,Faa-1857}\footnote{
For this formula one may also refer to \cite{Johnson-2002}.}
\begin{equation}\label{Y-p-Faa}
Y_{[nx]}(h)=\sum\frac{n!}{c_{1}!\cdots c_{n}!(1!)^{c_{1}}\cdots (n!)^{c_{n}}}h_{1}^{c_{1}}\cdots h_{n}^{c_{n}}, ~~~~~
			n=1,2,\cdots,
\end{equation}
where the sum is to be taken over all partitions of $ n= c_{1}+2c_{2}+\cdots +nc_{n}$.
The first few Bell polynomials are
\begin{equation*}
\begin{aligned}
			&Y_{[x]} =h_1,\\
			&Y_{[2x]} =h_2+h_{1}^{2},\\
			&Y_{[3x]} =h_3+3h_1h_2+h_{1}^{3}.
\end{aligned}
\end{equation*}

In the following, for convenience, we call the polynomials defined in \eqref{Y-def}
$Y$-polynomials.
Using the relation
\begin{equation}
(FG)^{-1}\partial^{n}_{x}(FG)
=\sum_{p=0}^{n}\binom{n}{p}(F^{-1}\partial^{n-p}_{x}F)(G^{-1}\partial^{p}_{x}G),
\end{equation}
and let $F={\mathrm{e}}^{h}$ , $G={\mathrm{e}}^{h^{\prime}}$,
one can obtain an addition formula for $Y$-polynomials:
\begin{align}
Y_{[nx]}(h+h^{\prime}) &={\mathrm{e}}^{-(h+h^{\prime})}\partial^{n}_{x}{\mathrm{e}}^{(h+h^{\prime})}
=({\mathrm{e}}^{-h} {\mathrm{e}}^{-h^{\prime}}) \partial^{n}_{x}({\mathrm{e}}^{h} {\mathrm{e}}^{h^{\prime}}) \nonumber \\
&=\sum_{p=0}^{n}\binom{n}{p} ({\mathrm{e}}^{-h}\partial^{n-p}_{x}{\mathrm{e}}^{h})
( {\mathrm{e}}^{-h^{\prime}} \partial^{p}_{x}{\mathrm{e}}^{h^{\prime}}) \nonumber \\
&=\sum_{p=0}^{n}\binom{n}{p}Y_{[(n-p)x]}(h)Y_{[px]}(h^{\prime}). \label{2.5}
\end{align}
In addition, replacing $h(x)$ by $h(-x)$ in \eqref{Y-def}, one can immediately find that
\begin{align}
Y_{n}(-h_{1},h_{2}, \cdots ,(-1)^{n}h_{n})
=(-1)^{n}Y_{[nx]}(h). \label{2.6}
\end{align}
This also indicates that
\begin{equation}\label{2.7}
Y_{[(2n+1)x]}(h)|_{h_1=h_3=\cdots=h_{2n+1}=0}=0.
\end{equation}

\subsection{Bilinear derivatives and Bell polynomials}\label{sec-2-2}

For two $\mathrm{C}^{\infty}$ functions $F(x)$ and $G(x)$,
their $N$th order bilinear derivative was introduced by Hirota in 1974 \cite{Hir-1974}:
\begin{equation}\label{D}
	D^n_x F \cdot G=(\partial_x-\partial_ {x'})^n F(x) G(x')|_{x'=x},
\end{equation}
where $D$ is called Hirota's bilinear operator.
Since
\begin{equation} 
	D^n_x F \cdot G=\sum^{n}_{p=0}(-1)^{p}\binom{n}{p}(\partial^{n-p}_{x}F)(\partial^{p}_{x}G),
\end{equation}
if we introduce
\[F ={\mathrm{e}}^{f(x)},~~~ G ={\mathrm{e}}^{g(x)},\]
it is easy to see that
\begin{align}
(FG)^{-1}D^{n}_{x} F \cdot G
&=\sum^{n}_{p=0}(-1)^{p}\binom{n}{p}(F^{-1}\partial^{n-p}_{x}F)(G^{-1}\partial^{p}_{x}G) \nonumber \\
&=\sum^{n}_{p=0}(-1)^{p}\binom{n}{p}Y_{[(n-p)x]}(f) Y_{[px]}(g),
\label{D-Y-1}
\end{align}
which is expressed in terms of $Y$-polynomials of $f$ and $g$.
Next, using the formulae \eqref{2.5} and \eqref{2.6},
the above relation can be written as
\begin{equation}\label{D-Y-2}
(FG)^{-1}D^{n}_{x} F\cdot{G}=
Y_{n}(h_{1},\cdots, h_{n})|_{h_{m}= f_{m}+(-1)^{m}g_{m}}.
\end{equation}
Equation \eqref{D-Y-1} and \eqref{D-Y-2} reveal the relations between Hirota's bilinear derivatives and
the Bell polynomials.

In practice, it is more convenient to use \eqref{D-Y-2} in its version described in terms of binary Bell polynomials:
\begin{equation}\label{Bell-binary}
\mathcal{Y}_{[nx]}(v,u)=Y_{n}(h_{1},\cdots,h_{n})|_{h_{2j-1}=v_{2j-1},\,h_{2j}=u_{2j}}.
\end{equation}
Note again that for function $f(x)$, by $f_k$ we denote $\partial^k_x f(x)$.
For convenience, we call $\mathcal{Y}_{[nx]}(v,u)$ $\mathcal{Y}$-polynomials.
The first few of them read
\begin{subequations}
\begin{align}
		&\mathcal{Y}_{[x]}(v,u)=v_{1},\\
		&\mathcal{Y}_{[2x]}(v,u)=u_{2}+v^{2}_{1},\\
		&\mathcal{Y}_{[3x]}(v,u)=v_{3}+3u_{2}v_{1}+v^{3}_{1},\\
		&\mathcal{Y}_{[4x]}(v,u)=u_{4}+4v_{3}v_{1}+3u^{2}_{2}+6u_{2}v^{2}_{1}+v^{4}_{1}.
\end{align}
\end{subequations}

Thus, the relation \eqref{D-Y-2} is written as
\begin{equation}\label{D-Y-3}
	(FG)^{-1}D^{n}_{x} F\cdot G =\mathcal{Y}_{[nx]}(v=\ln (F/G), u=\ln (FG)),
\end{equation}
where we have taken $u=f+g$ and $v=f-g$.
In particular, the bilinear derivatives $D^n_x F\cdot F$ can be expressed using $\mathcal{Y}_{[nx]}(0,u)$.
When $n$ is odd, from the definition \eqref{Bell-binary} together with property \eqref{2.7}, we have
\begin{equation}\label{Y-odd}
	\mathcal{Y}_{[(2j+1)x]}(v=0,u) =0,
\end{equation}
which agrees with the property $D^{2j+1}_x F\cdot F=0$.
If $n$ is even, we denote ($P$-polynomials for short)
\begin{equation}\label{P}
		P_{[2jx]}(u) \equiv \mathcal{Y}_{[2jx]}(0,u),
\end{equation}
the first few of which are
\begin{subequations}\label{P-polys}
	\begin{align}
		&P_{[0]}(u) = 1,\\
		&P_{[2x]}(u) = u_{2},\\
		&P_{[4x]}(u) = u_{4}+3u^{2}_{2},\\
		&P_{[6x]}(u) = u_{6}+15u_{2}u_{4}+15u_{2}^{3}.
	\end{align}
\end{subequations}
Thus, for $n=2j$ being even, we have
\begin{equation}
	F^{-2}D^{2j}_{x} F\cdot F =P_{[2jx]}(u=2\ln F).
\end{equation}

Note that the binary Bell polynomials  $\mathcal{Y}_{[nx]}(v,u)$
can be expressed using $Y$-polynomials and $P$-polynomials.
In fact,
\begin{align*}
\mathcal{Y}_{[nx]}(v,u) &= \mathcal{Y}_{[nx]}(v,v+u-v)
= Y_{n}(v_{1}+0,v_{2}+(u-v)_{2},\cdots )\\
& = Y_{[nx]}(v+h)|_{h_{2j-1}=0, h_{2j}=(u-v)_{2j}}\\
&= \sum_{r=0}^{n}\binom{n}{r}Y_{[(n-r)x]}(v)Y_{[rx]}(h)|_{h_{2j-1}=0, h_{2j}=(u-v)_{2j}},
\end{align*}
where the addition formula \eqref{2.5} has been used.
Then, using \eqref{Bell-binary}, \eqref{Y-odd} and \eqref{P},
we have
\begin{equation}
\mathcal{Y}_{[nx]}(v,u)
=\sum^{[n/2]}_{j=0}\binom{n}{2j}Y_{[(n-2j)x]}(v)P_{[2jx]}(u-v),
\end{equation}
where $[n/2]$ denotes the floor function of $n/2$.

\subsection{Multidimensional generalization}\label{sec-2-3}

Assume $h=h(x_1,x_2,\cdots, x_{\ell})$ is an $\ell$-dimension C$^{\infty}$ function of $(x_1,x_2,\cdots, x_{\ell})$.
Denote
\begin{equation*}
h_{r_{1},\cdots,r_{\ell}} =\partial^{r_{1}}_{x_{1}}\cdots\partial^{r_{\ell}}_{x_{\ell}}h(x_{1},\cdots,x_{\ell}).
\end{equation*}
The $\ell$-dimension Bell polynomials of $h$ are defined by
\begin{equation}\label{Bell-el}
Y_{[n_{1}x_{1}, \cdots, n_{\ell}x_{\ell}]}(h) \doteq
Y_{n_{1}, \cdots, n_{\ell}}(\{h_{r_{1}, \cdots, r_{\ell}}\})
={\mathrm{e}}^{-h}\partial^{n_{1}}_{x_{1}} \cdots \partial^{n_{\ell}}_{x_{\ell}}{\mathrm{e}}^{h},
\end{equation}
which are alteratively expressed through
\begin{equation}
Y_{[n_{1}x_{1}, \cdots,n_{\ell}x_{\ell}]}(h)
=\sum\frac{n_{1}!n_{2}! \cdots n_{\ell}!}{c_{1}!c_{2}! \cdots c_{k}!}
\prod^{k}_{j=1}\left(\frac{h_{r_{1j}, \cdots,r_{\ell\textit{j}}}}{r_{1j}! \cdots r_{\ell\textit{j}}!}\right)^{c_{j}}.
\end{equation}
Here we just give two examples. When $h=h(x_1,x_2)$, we have
\begin{align*}
& Y_{[x_1,x_2]}=h_{1,0}h_{0,1}+h_{1,1},\\
& Y_{[2x_1,x_2]}=2h_{1,0}h_{1,1}+h_{1,0}^2 h_{0,1}+h_{2,0}h_{0,1}+h_{2,1},\\
& Y_{[3x_1,x_2]}=3h_{1,0}^2h_{1,1}+3h_{1,1} h_{2,0}+3h_{1,0}h_{2,1}
+h_{0,1}(h_{1,0}^3+3h_{1,0}h_{2,0}+h_{3,0})+h_{3,1},
\end{align*}
and when $h=h(x_1,x_2,x_3)$, we have
\[Y_{[x_1,x_2,x_3]}=h_{1,0,0}h_{0,1,1}+h_{1,0,1}h_{0,1,0}+h_{1,1,0}h_{0,0,1}
+h_{1,0,0}h_{0,1,0}h_{0,0,1}+h_{1,1,1}.\]
These polynomials allow addition formula
\begin{equation}
Y_{[n_{1}x_{1}, \cdots, n_{\ell}x_{\ell}]}(f+g)
=\sum_{p_{1}=0}^{n_{1}}\cdots \sum_{p_{\ell}=0}^{n_{\ell}}
\prod_{i=1}^{\ell}\binom{n_{i}}{p_{i}}Y_{[(n_{1}-p_{1})x_{1}, \cdots, (n_{\ell}-p_{\ell})x_{\ell}]}(f)
Y_{[p_{1}x_{1}, \cdots, p_{\ell}x_{\ell}]}(g)
\end{equation}
and an identity
\begin{equation}
Y_{n_{1}, \cdots, n_{\ell}}(\{(-1)^{r_{1}+ \cdots +r_{\ell}}h_{r_{1}, \cdots, r_{\ell}}\})
=(-1)^{n_{1}+ \cdots +n_{\ell}}Y_{n_{1}, \cdots, n_{\ell}}(\{h_{r_{1}, \cdots, r_{\ell}}\}).
\end{equation}

For $F$ and $G$ being  C$^{\infty}$ functions of $\mathbf{x}=(x_1,x_2,\cdots, x_{\ell})$,
their bilinear derivative is defined as
$$ D^{n_{1}}_{x_{1}}\cdots D^{n_{\ell}}_{x_{\ell}} F(\mathbf{x})\cdot G(\mathbf{x}) =(\partial_{x_{1}}-\partial_{x_{1}^{\prime}})^{n_{1}}\cdots
(\partial_{x_{\ell}}-\partial_{x_{\ell}^{\prime}})^{n_{\ell}}
F(\mathbf{x}) G(\mathbf{x}^\prime)|_{\mathbf{x}^{\prime}=\mathbf{x}}.$$
Introducing  $F = {\mathrm{e}}^{f(\mathbf{x})} $ and $G ={\mathrm{e}}^{g(\mathbf{x})}$, we can have
\begin{equation}\label{D-Y-5}
(FG)^{-1}D_{x_{1}}^{n_{1}}\cdots D_{x_{\ell}}^{n_{\ell}}F \cdot G
=Y_{n_{1},\cdots,n_{\ell}}(\{h_{r_{1},\cdots,r_{\ell}}=f_{r_{1},\cdots,r_{\ell}}+(-1)^{r_{1}
+\cdots+r_{\ell}}g_{r_{1},\cdots, r_{\ell}}\}).
\end{equation}

We now introduce multidimensional binary Bell polynomials \cite{gilson-1996,lambert-1994}:
\begin{subequations}\label{Bell-binary-md}
\begin{equation}
\mathcal{Y}_{[n_{1}x_{1},\cdots,n_{\ell}x_{\ell}]}(v,u) \doteq
Y_{n_{1},\cdots,n_{\ell}}(\{h_{r_{1},...,r_{\ell}}\})
\end{equation}
where
\begin{equation}
h_{r_{1},\cdots,r_{\ell}}=\biggl\{\begin{array}{ll}
u_{r_{1},\cdots,r_{\ell}}, & \mathrm{if}~ r_{1}+\cdots+r_{\ell}  ~\mathrm{is~ even}, \\
v_{r_{1},\cdots,r_{\ell}}, & \mathrm{if }~ r_{1}+\cdots+r_{\ell}  ~\mathrm{is~ odd}.
\end{array}
\end{equation}
\end{subequations}
Similar to the 1-dimension case, we have
\[\mathcal{Y}_{[n_{1}x_{1},\cdots,n_{\ell}x_{\ell}]}(v=0,u)=0,~~~
\mathrm{if}~ n_{1}+\cdot\cdot\cdot+n_{\ell}~ \mathrm{is~ odd}.\]
When $n_{1}+\cdots+n_{\ell}$ is even, we denote
\begin{align}
P_{[n_{1}x_{1},\cdots,n_{\ell}x_{\ell}]}(u) =\mathcal{Y}_{[n_{1}x_{1},\cdots,n_{\ell}x_{\ell}]}(v=0,u).
\end{align}
In 2-dimension case, i.e., $\ell=2$ and $x_1=x, x_2=y$, the first few $\mathcal{Y}$-polynomials are
\begin{subequations}
\begin{align}
	&\mathcal{Y}_{[x]}(v,u) =v_{x},\\
	&\mathcal{Y}_{[2x]}(v,u) =u_{2x}+v^{2}_{x},\\
	&\mathcal{Y}_{[x,y]}(v,u) =u_{x,y}+v_{x}v_{y},\\
	&\mathcal{Y}_{[3x]}(v,u) =v_{3x}+3v_{x}u_{2x}+v_{x}^{3},\\
	&\mathcal{Y}_{[2x,y]}(v,u) =v_{2x,y}+2v_{x}u_{x,y}+v^{2}_{x}v_{y}+u_{2x}v_{y},\\
    &\mathcal{Y}_{[3x,y]}(v,u) =3v^2_x u_{x,y}+3u_{x,y}u_{2x}+3 v_{x}v_{2x,y}
       +v_{y}(v^3_{x}+3v_x u_{2x}+v_{3x})+ u_{3x,y}.
\end{align}
\end{subequations}
The simplest example of 3-dimension case ($(x_1,x_2,x_3)=(x,y,z)$) is
\begin{equation}
  \mathcal{Y}_{[x,y,z]}(v,u) =v_{x,y,z}+v_{x}u_{y,z}+v_{y}u_{x,z}+v_{z}u_{x,y}+v_{x}v_{y}v_{z}.
\end{equation}
The first few $P$-polynomials are
\begin{subequations}
\begin{align}
	& P_{[2x]}(u) =u_{2x},\\
	& P_{[x,y]}(u) =u_{x,y},\\
    & P_{[2x,2y]}(u) =2u^2_{x,y}+u_{2x} u_{2y}+u_{2x,2y},\\
	& P_{[3x,y]}(u) =u_{3x,y}+3u_{2x}u_{x,y}.
\end{align}
\end{subequations}
Here, we use $u_{kx,sy}$ to denote $\partial_x^k\partial_y^s u(x,y)$ without making any confusion.

With regard to the connection between the multidimensional binary Bell polynomials and
bilinear derivatives of $F$ and $G$,
by taking $F ={\mathrm{e}}^{f}, G ={\mathrm{e}}^{g}$,  $ u = f+g$ and $v=f-g$, we can have (cf.\eqref{D-Y-5})
\begin{equation}\label{2.29}
(FG)^{-1}D^{n_{1}}_{x_{1}}\cdots D^{n_{\ell}}_{x_{\ell}} F\cdot G
=\mathcal{Y}_{[n_{1}x_{1},...,n_{\ell}x_{\ell}]}(v=\ln(F/G), u=\ln(FG)),
\end{equation}
i.e.,
\begin{equation}\label{D-Y-6}
(FG)^{-1}D^{n_{1}}_{x_{1}}\cdots D^{n_{\ell}}_{x_{\ell}} F \cdot G
=\sum_{p_{1}=0}^{n_{1}}\cdots \sum_{p_{\ell}=0}^{n_{\ell}}
\prod_{i=1}^{\ell}\binom{n_{i}}{p_{i}}
Y_{[(n_{1}-p_{1})x_{1},\cdots,(n_{\ell}-p_{\ell})x_{\ell}]}(v)P_{[p_{1}x_{1},\cdots,p_{\ell}x_{\ell}]}(2g),
\end{equation}
and in particular, when $F=G$, we have
\begin{align}\label{2.30}
F^{-2}D^{n_{1}}_{x_{1}}\cdots  D^{n_{\ell}}_{x_{\ell}}F \cdot F
= P_{[n_{1}x_{1},...,n_{\ell}x_{\ell}]}(u=2\ln F)
\end{align}
where $n_{1}+\cdots +n_{\ell}$ is even.
These formulae provide relations that will be used to convert Hirota's bilinear equations into nonlinear forms.

Finally, we also point out that, from the definition \eqref{Bell-el},
any Bell polynomial define in \eqref{Bell-el} can be linearized through a logarithmic transformation
$h(\mathbf{x})=\ln \psi(\mathbf{x})$:
\begin{equation}\label{hct}
  \psi\,  Y_{[n_{1}x_{1},\cdots,n_{\ell}x_{\ell}]}(h)|_{h=\ln\psi}
  = \psi_{n_{1}x_{1},\cdots,n_{\ell}x_{\ell}}.
\end{equation}
This is useful in deriving linear problems (Lax pairs)
from bilinear B\"acklund transformations, e.g.\cite{Lambert2001}.

\section{Convert Hirota bilinear equations to nonlinear forms}\label{sec-3}

In this section, we provide examples to show how nonlinear equations are recovered from their bilinear forms.

\subsection{Examples of the KdV-type}\label{sec-3-1}

The KdV-type bilinear equations are \cite{Hir-1980}
\begin{equation}\label{KdV-type}
Q(D_{x},D_{y},D_t,\cdots)F\cdot F=0,
\end{equation}
where $Q$ is a polynomial with constant coefficients and $F=0$ must be a solution of \eqref{KdV-type}.
In 1980, Hirota showed that a KdV-type bilinear equation always has 1-soliton and 2-soliton solutions
and he proposed 3-soliton condition as an integrable criteria for bilinear equations of the KdV type \cite{Hir-1980}.
Later, Hietarinta searched the KdV-type bilinear equations that have 3-soliton solutions \cite{Hie-1987a}.
Some examples on the Hietarinta's list are (see Table IV in \cite{Hie-1987a})
\begin{subequations}\label{Hietarinta-list}
\begin{align}
(1)~~ & (D_{x}^{4}-4D_{x}D_{t}+3D_{y}^{2})F\cdot F=0, \label{KP-b}\\
(2)~~ & (D_{x}^{3}D_{t}+aD_{x}^{2}+D_{t}D_{y})F\cdot F=0, \label{HS-a}\\
(3)~~ & (D_{x}^{6}+5D_{x}^{3}D_{t}-5D_{t}^{2}+D_{x}D_{y})F\cdot F=0, \label{BKP}\\
(4)~~ & [D_{x}D_{t}(D_{x}^{2}+\sqrt{3}D_{x}D_{t}+D_{t}^{2})+aD_{x}^{2}+bD_{x}D_{t}+cD_{t}^{2}]
F\cdot F=0,
\label{Hie-eq}
\end{align}
\end{subequations}
where $a,b,c$ are constants.
Among them, \eqref{KP-b} is the well known bilinear KP equation.
\eqref{HS-a} and \eqref{BKP} belongs to the bilinear BKP hierarchy
(see page 768 and 769 in \cite{DJM-JPSJ-1983})
and allows various reductions (see \cite{Hir-JPSJ-1989} and the references therein).
\eqref{Hie-eq} is new, now called Hietarinta's equation (cf.\cite{ZHS-JCP-2018}),
and its nonlinear form has not yet been known.
The first three equations allow dimension reductions of traveling wave type, e.g. $D_y=a_1 D_x+a_2 D_y$
and the resulting 2-dimensional equations still admit 3-soliton solutions \cite{Hie-1987a}.
Apart from the above equations, all 2-dimensional equations in the ``Accepted final result'' columns in Table I, II and III
in \cite{Hie-1987a} provide  1-dimensional  and 2-dimensional bilinear equations that have 3-soliton solutions.

In what follows, we show how bilinear equations are converted into nonlinear forms by using the formulae
we got in Sec.\ref{sec-2}.
First, for the bilinear KP equation \eqref{KP-b}, taking
\begin{equation}
q=2\ln F
\label{q-F}
\end{equation}
and using formula \eqref{2.30}, we have
\begin{align}
& F^{-2}(D_{x}^{4}-4D_{x}D_{t}+3D_{y}^{2})F\cdot F \nonumber\\
=\,&P_{[4x]}(q)-4P_{[x,t]}(q)+3P_{[2y]}(q) \nonumber\\
=\,& q_{4x}+3q^{2}_{2x}-4q_{x,t}+3q_{2y}.
\label{3.4}
\end{align}
Letting  $u=q_{2x}$, we get the well known KP equation
\begin{equation}\label{KP-eq}
		(u_{xxx}+6uu_{x}-4u_{t})_{x}+3u_{yy}=0.
\end{equation}

For the second equation in \eqref{Hietarinta-list}, using \eqref{q-F} and \eqref{2.30}, the equation \eqref{HS-a} gives to
\begin{equation}\label{3.6}
P_{[3x,t]}(q)+a P_{[2x]}(q) + P_{[t,y]}(q)=q_{3x,t}+3q_{2x}q_{x,t}+aq_{2x}+q_{t,y}=0.
\end{equation}
Setting $u=q_{x}$ we get
\begin{equation}\label{u-BLMP}
	u_{xxxt}+au_{xx} + 3 (u_x u_t)_x +u_{yt} =0,
\end{equation}
which was first derived (together with the bilinear form \eqref{HS-a}) in \cite{DJM-JPSJ-1983}
as a member in the BKP hierarchy.\footnote{This equation is given in item (4)
in page 699 of \cite{DJM-JPSJ-1983},
corresponding a plane wave factor ${\mathrm{e}}^{kx+k^3 y+ k^{-1} t}$ up to some scaling.
It can be considered as the first member in the negative BKP hierarchy.
So, strictly speaking, it should be called the negative BKP equation or BKP$(-1)$ equation for short.
}
It is also a (2+1)-dimensional extension of the model equation for shallow water waves ($y=x$)
(cf. equation (2) and (7) \cite{HS-JPSJ-1976}).
In addition, introducing $w$ and $v$ by
\begin{equation}
w=u_t, ~~ v=u_x,
\label{wuv-BKP}
\end{equation}
equation \eqref{u-BLMP} is written as
\begin{equation}\label{BLMP}
	w_y+w_{xxx}+av_{x} + 3 (w v)_x =0, ~~~ w_x=v_t.
\end{equation}
This is known as the Boiti-Leon-Manna-Pempinelli (BLMP) equation (see equation (2.16) in \cite{BLMP-IP-1986}),
which is connected with the bilinear form \eqref{HS-a} by $w=2(\ln F)_{xt}$ and $v=2(\ln F)_{xx}$.
Of course, it is actually the BKP$(-1)$ equation \eqref{u-BLMP} via \eqref{wuv-BKP}.
Note that the term $av_x$ in this equation can be removed by a simple transformation
$w=\bar{w}-\frac{a}{3}$. Thus, on the nonlinear level, the parameter $a$ in the bilinear equation \eqref{HS-a}
maybe not important.

For the third equation \eqref{BKP}, by using  \eqref{q-F} and \eqref{2.30}, it yields
\begin{equation}
q_{6x}+15q_{2x}q_{4x}+15q_{2x}^{3}+5(q_{3x,t}+3q_{2x}q_{x,t})-5q_{2t}+q_{x,y}=0,
\end{equation}
which, by setting $u=q_{x}$, leads to
\begin{equation}
		(u_{y}+15u_{x}u_{xxx}+15u_{x}^{3}+15u_{x}u_{t}+u_{xxxxx})_{x}+5u_{xxxt}-5u_{tt}=0,
\end{equation}
which is the BKP equation (see item (2) on page 768 in \cite{DJM-JPSJ-1983},
corresponding to the plane wave factor ${\mathrm{e}}^{kx+k^3 t+ k^{5} y}$ up to some scaling).

For the Hietarinta equation \eqref{Hie-eq}, from \eqref{q-F} and \eqref{2.30} we get
\begin{equation}
q_{3x,t}+3q_{2x}q_{x,t}+\sqrt{3}(q_{2x,2t}+q_{2x}q_{2t}+2q_{x,t}^{2})
+q_{x,3t}+3q_{2t}q_{x,t}+aq_{2x}+bq_{x,t}+cq_{2t}=0.
\end{equation}
We can either use it as the nonlinear form of  \eqref{Hie-eq}, or after introducing $u=q_x$, get
\begin{equation}\label{Hie-eq-n}
u_{xxt}+u_{ttt}+3u_{t}(u_{x}+\partial_x^{-1}u_{tt})
+\sqrt{3}(u_{xtt}+u_{x}\partial_x^{-1}u_{tt}+2u_{t}^{2})+au_{x}+bu_{t}+c\partial_x^{-1}u_{tt}=0,
\end{equation}
where $\partial_x^{-1}\cdot=\int^x_{-\infty}\cdot\, dx$.
A third nonlinear form of \eqref{Hie-eq} is (by introducing $u=q_x,~ v=q_t$)
\begin{equation}\label{Hie-eq-n-uv}
v_{xxx}+v_{xtt}+3v_{x}(u_{x}+ v_{t})
+\sqrt{3}(v_{xxt}+u_{x}v_{t}+2v_{x}^{2})+au_{x}+bv_{x}+cv_{t}=0, ~~ v_x=u_t.
\end{equation}

All the equations in Hietarinta's list \eqref{Hietarinta-list}
allow traveling-wave extension, in other words,
e.g. replacing $D_y$ by $D_y+D_z$ in \eqref{KP-b}, the obtained bilinear equation still has 3-soliton solution.
For the bilinear KP equation \eqref{KP-b}, such an extension yields
\begin{equation}
(D_{x}^{4}-4D_{x}D_{t}+3(D_{y}+D_{z})^{2})F\cdot F=0.
\end{equation}
The resulting nonlinear equation (via $q=2\ln F$ and $u=q_{2x}$)
reads	
\begin{equation}\label{KP-ed}
		(u_{xxx}+6uu_{x}-4u_{t})_{x}+3(u_{yy}+2u_{yz}+u_{zz})=0.
\end{equation}
Such type of extended KP equations have been frequently studied, e.g. \cite{MWL-2021,MZM-2018}.
However, the travelling-wave type extension is trivial since solution of the extended equation
can be easily obtained by making corresponding replacement of independent variables in the
elementary plane wave factor,
e.g., for \eqref{KP-ed}, just replacing $y\to y+z$ in the plane wave factor of the KP solutions.

One more integrable example is a 4-dimensional bilinear equation
\begin{equation}\label{KP-4D}
(D_{x}^{3}D_y + 2 D_{y}D_{t}-3D_{x}D_{z})F\cdot F=0,
\end{equation}
which appears in the bilinear KP hierarchy (see page 995 of \cite{JimboM-1983} and equation (4.36b)
in \cite{LZ-JNS-2022}), but not the BKP hierarchy.
However, it will recover the bilinear BKP$(-1)$ equation \eqref{HS-a} by taking $z=x$.
It is interesting that the bilinear equation is also related to toroidal algebra
$\mathrm{sl}_2^{\mathrm{tor}}$ (see (3.6) in \cite{Billig-JA-1999} 
and degree 3 case in Table 1 in \cite{IT-IMRN-2001}, up to some scaling).
Note that the bilinear equation \eqref{KP-4D} coupled with the bilinear KdV equation
$(D_x^4-4D_xD_t)F\cdot F=0$ provide a bilinear form for Bogoyavlensky's
breaking soliton KdV equation \cite{IT-IMRN-2001}.
Using $q=2\ln F$ and  formula \eqref{2.30}, from the above bilinear equation we have
\begin{align*}
P_{[3x,y]}(q)+2P_{[y,t]}(q)-3P_{[x,z]}(q)
  =q_{3x,y}+3q_{2x}q_{x,y}+2q_{y,t}-3q_{x,z}=0.
\end{align*}
Thus, we obtain a nonlinear form of \eqref{KP-4D}:
(via $q=2\ln F$ and $u=q_{x}$)
\begin{equation}\label{KP-4D-u}
    u_{xxxy}+3(u_{x}u_{y})_{x}+2u_{yt}-3u_{xz}=0.
\end{equation}
If we further introduce $w=u_y$ and $v=u_x$, we get
\begin{equation}\label{KP-4D-w}
    w_{xxx}+3(wv)_{x}+2w_{t}-3v_{z}=0, ~~~ v_y=w_x,
\end{equation}
which is now connected with the bilinear equation \eqref{KP-4D} via
$w=2(\ln F)_{xy}$ and $v=2(\ln F)_{xx}$.
Compared with the BLMP equation \eqref{BLMP} (switching $t\leftrightarrow y$),
we may consider \eqref{KP-4D-w} as a 4 dimensional extension  of \eqref{BLMP},
or, we say \eqref{KP-4D-u} is a 4 dimensional extension of the BKP$(-1)$ equation \eqref{u-BLMP}.

\subsection{Non-integrable examples of the KdV-type}\label{sec-3-2}

As we mentioned, Hirota showed that a KdV-type bilinear equation always has
1-soliton and 2-soliton solutions \cite{Hir-1980}
but may not have a 3-soliton solution.
These means some nonlinear PDEs might not be integrable although they have 2-soliton solutions.
In the following we provide three such non-integrable examples of the KdV type.

The first example is
\begin{equation}\label{n-KdV-1}
(D_{x}^{3}D_{y}+D_{x}D_{t})F\cdot F =0.
\end{equation}
Note that this equation is not integrable in Hirota's sense because it does not have a 3-soliton solution.\footnote{
In fact, this equation has the following solutions
\begin{align*}
& F=1+ {\mathrm{e}}^{\xi_1},~~ F=1+{\mathrm{e}}^{\xi_1}+{\mathrm{e}}^{\xi_2}+A_{12}{\mathrm{e}}^{\xi_1+\xi_2},\\
& \xi_i=k_ix+s_i y -s_ik_i^2 t +\xi_i^{(0)},~~
A_{ij}=\frac{(k_i-k_j)[s_j k_i (k_i - 2 k_j) + s_i (2 k_i - k_j) k_j]}
{(k_i+k_j)[s_j k_i (k_i + 2 k_j) + s_i (2 k_i + k_j) k_j]},
\end{align*}
where $k_i, s_i$ and $\xi_i^{(0)}$ are constants.
However, it can be checked that
\begin{equation*}
F=1+{\mathrm{e}}^{\xi_1}+{\mathrm{e}}^{\xi_2}+{\mathrm{e}}^{\xi_3}+A_{12}{\mathrm{e}}^{\xi_1+\xi_2}+A_{13}{\mathrm{e}}^{\xi_1+\xi_3}+A_{23}{\mathrm{e}}^{\xi_2+\xi_3}
+A_{12}A_{13}A_{23}{\mathrm{e}}^{\xi_1+\xi_2+\xi_3}
\end{equation*}
is not a solution of \eqref{n-KdV-1}.
}
By setting  $q=2\ln F$ and using formula \eqref{2.30}, we obtain
\begin{align}
 P_{3x,y}(q)+P_{x,t}(q)= q_{xxxy}+3q_{xx}q_{xy}+q_{xt}=0.
\end{align}
If we just take $w=q_x$, we have a simple equation
\begin{equation}
w_t+w_{xxy}+3 w_{x}w_y=0.
\end{equation}
If we replace $q$ by $u=q_{2x}$, we get
\begin{equation}\label{u-kdv-n}
		u_{t}+3uu_{y}+u_{xxy}+3u_{x}\int_{-\infty}^{x}u_{y} \,dx=0.
\end{equation}
The above equation \eqref{u-kdv-n} has been studied in \cite{zhang-2006,MZG-MPLA-2009}.
Unfortunately, it was considered to be integrable in \cite{ZCDD-CTP-2011}.

The second example is
\begin{equation}
		(D_{t}D_{y}-D_{x}^{3}D_{y}-3D_{x}^{2}+3D_{z}^{2})F\cdot F=0,
\end{equation}
which was first considered as an example allows resonance of solitons in \cite{GZZYL-2016}.
Setting  $q=2\ln F$ and using formula \eqref{2.30}, and then taking  $u=q_{x}$,
one can get its nonlinear form
\begin{equation}
		u_{ty}-u_{xxxy}-3(u_{x}u_{y})_{x}-3u_{xx}+3u_{zz}=0.
\end{equation}
Note that this is again  not integrable (see \cite{liu-2021}),
although sometimes it was studied incorrectly as an integrable equation, e.g. \cite{DTYZ-2018}.

Since the KdV-type bilinear equations always admit 1-soliton and 2-soliton solutions, even they are not integrable,
one can freely choose a KdV-type bilinear equation, then the corresponding nonlinear form
will always have 1-soliton and 2-soliton solutions.  Here we just add one more non-integrable example,
which is related to the bilinear form
\begin{equation}
(D_xD_t+D^2_xD_y^2)F\cdot F=0,
\end{equation}
and the nonlinear form reads (via $w(x,y,t)=2(\ln F)_x$)
\begin{equation}
w_t+2w_y^2+w_{xyy}+w_x\int_{-\infty}^{x}w_{yy}\,dx=0.
\end{equation}

\subsection{Examples of the mKdV-type}\label{sec-3-3}

In addition to the KdV-type bilinear equations, the mKdV-type and the sG-type bilinear equations
automatically admit 1-soliton and  2-soliton solutions as well \cite{Hie-1987b,Hie-1987c}.
Note that the NLS-type bilinear equations considered by Hietarinta always have 1-soliton solutions \cite{Hie-1988}.
Apart from the KdV-type bilinear equations, Hietarinta also did the searching of the
mKdV-type and sG-type bilinear equations that have 3-soliton solutions \cite{Hie-1987b,Hie-1987c}.

In the following, we provide some examples of the mKdV-type bilinear equations,
of which the general form (in 2-dimension as an example) considered in \cite{Hie-1987b} is a coupled system
\begin{subequations}
\begin{align}
& Q_1(D_x,D_t) F\cdot G=0, \\
& Q_2(D_x,D_t) F\cdot G=0,
\end{align}
\end{subequations}
where $Q_1$ and $Q_2$ are some 2-component polynomials,
$Q_1$ is odd and $Q_2$ is quadratic and even.
To convert them into nonlinear equations, we need to use formula \eqref{2.29} or \eqref{D-Y-6},
and the transformations
\begin{equation}\label{trans-mkdv}
u=\ln(FG),~~ v=\ln (F/G).
\end{equation}

The first example of the mKdV-type is
\begin{subequations}\label{5.12}
	\begin{align}
		&(D_{t}+D_{x}^{3})F\cdot G=0 \label{5.12a},\\
		&D_{x}^{2}F\cdot G=0  \label{5.12b}.
	\end{align}
\end{subequations}
Based on formula \eqref{2.29}, we can derive
\begin{align}
(D_{t}+D_{x}^{3})F\cdot G
&=\mathcal{Y}_{[t]}(v,u)+\mathcal{Y}_{[3x]}(v,u) \nonumber \\
&=v_{t}+v_{xxx}+3v_{x}u_{xx}+v_{x}^{3}=0.
\label{5.13}
\end{align}
Meanwhile, from equation \eqref{5.12b}, we have
\begin{align}
		D_{x}^{2}F\cdot G=\mathcal{Y}_{[2x]}(v,u)=u_{xx}+v_{x}^{2}=0. \label{5.14}
\end{align}
Substituting the above equation into \eqref{5.13} to eliminate $u_{xx}$,
we arrive at the potential (defocusing) mKdV equation
\begin{equation}
v_{t}+v_{xxx}-2v_{x}^{3}=0,
\end{equation}
which yields the mKdV equation ($w=v_x=\partial_x\ln(f/G)$)
\begin{equation}
		w_{t}+w_{xxx}-6w^{2}w_{x}=0.
\end{equation}

Another  example of the mKdV-type is (see Table I (Generalizations with $Y$) of \cite{Hie-1987b})
\begin{subequations}\label{5.17}
\begin{align}
		&(D_{x}^{2}D_{t}+D_{y})F \cdot G=0, \label{5.17a}\\
		&D_{x}^{2}F\cdot G=0.   \label{5.17b}
\end{align}
\end{subequations}
These equations were also derived in \cite{YC-LMP-2024}
as bilinear forms related to the  Bogoyavlensky-mKdV hierarchy  and 
toroidal Lie algebra $\mathrm{sl}^{\mathrm{tor}}_2$,
see degree 2 case in Table 1 in \cite{YC-LMP-2024}.
Using formula \eqref{2.29} we have
\begin{align}
(D_{x}^{2}D_{t}+D_{y})F \cdot G &=\mathcal{Y}_{[2x,t]}(v,u)+\mathcal{Y}_{[y]}(v,u) \nonumber \\
	&=v_{xxt}+2v_{x}u_{xt}+v_{x}^{2}v_{t}+u_{xx}v_{t}+v_{y}=0
\label{5.18}
\end{align}
and \eqref{5.14}. Eliminating $u$ we obtain
\begin{equation}
		v_{xxt}-4v_{x}\int_{-\infty}^{x} v_{x}v_{xt}\,dx+v_{y}=0,
\end{equation}
which leads us to (with $w= v_{x}=\partial_x\ln(F/G)$):
\begin{equation}\label{mkdv-b-3}
		w_{y}+w_{xxt}-4w^{2}w_{t}-2w_{x}\int_{-\infty}^{x}(w^2)_{t}\,dx=0.
\end{equation}
This is an integrable equation known as the breaking soliton mKdV equation
(see equation (1.4) on page 47 of \cite{Bogo-1990}).
Lax pair of this equation has been constructed from a generic sense in \cite{YC-LMP-2024}.
In addition, its alternative bilinear form has been given in \cite{Ohta-2003},
where apart from \eqref{5.17},  equation \eqref{5.12a} is also involved.

Next  example  is (also see Table I (Generalizations with $Y$) of \cite{Hie-1987b})
\begin{subequations}\label{5.177}
\begin{align}
		&(D_x^5+ D_{x}^{2}D_{t}+D_{y})F \cdot G=0, \label{5.177a}\\
		&D_{x}^{2}F\cdot G=0.   \label{5.177b}
\end{align}
\end{subequations}
The related Bell polynomial expressions are
\begin{align*}
  \mathcal{Y}_{[5x]}(v,u)+\mathcal{Y}_{[2x,t]}(v,u)+\mathcal{Y}_{[y]}(v,u)=0
\end{align*}
and \eqref{5.14},
which are given in terms of $(v,u)$ as
\begin{align*}
    &v_{5x}+v_{x}^{5}+10v_{x}^{2}v_{3x}+5v_{x}u_{4x}+10v_{3x}u_{2x}+15v_{x}u_{2x}^{2}\\
    &~~~~ +10v_{x}^{3}u_{2x}+v_{2x,t}+2v_{x}u_{x,t}+v_{x}^{2}v_{t}+u_{2x}v_{t}+v_{y}=0
\end{align*}
and $u_{2x}+v_{x}^{2}=0$.
Again, eliminating $u$ and introducing $w=v_{x}$, we arrive at
\begin{equation}\label{mkdv-b-5}
    w_{xxxxx}+30w^{4}w_{x}+w_{xxt}+w_{y}
    -\Bigl(10w^{2}w_{xx}+10ww_{x}^{2}+2w\int_{-\infty}^{x}(w^2)_{t}\,dx\Bigr)_{x}=0.
\end{equation}
It can be considered as a 5th-order  breaking soliton mKdV equation, cf.\eqref{mkdv-b-3},
and it is connected with the bilinear form \eqref{5.177} via $w= v_{x}=\partial_x\ln(F/G)$.

The last example is (see equation (34) in \cite{Hie-1987b})
\begin{subequations}\label{3.32}
\begin{align}
		&(D_{x}D_yD_{t}+a D_{x} + b D_t) F \cdot G=0, \label{3.32a}\\
		&D_{x}D_t F\cdot G=0,   \label{3.32b}
\end{align}
\end{subequations}
where $a, b$ are parameters.
It corresponds to
\begin{align*}
    &\mathcal{Y}_{[x,y,t]}(v,u)+a\mathcal{Y}_{[x]}(v,u)+b\mathcal{Y}_{[t]}(v,u)=0,\\
    &\mathcal{Y}_{[x,t]}(v,u)=0,
\end{align*}
i.e.
\begin{align*}
    &v_{xyt}+v_{x}u_{yt}+v_{y}u_{xt}+v_{t}u_{xy}+v_{x}v_{y}v_{t}+av_{x}+bv_{t}=0,\\
    &u_{xt}+v_{x}v_{t}=0.
\end{align*}
By introducing $w=u_y$, we have the following nonlinear form:
\begin{equation}\label{mkdv-xyt}
v_{xyt}+v_{x}w_{t}+v_{t}w_{x}+av_{x}+bv_{t}=0,~~~ w_{xt}=(v_xv_t)_y,
\end{equation}
which is connected with the bilinear equation \eqref{3.32} via
$v=\ln (F/G)$ and $w=\partial_y \ln (FG)$.
This equation is integrable  in Hirota's sense, while all its integrability characteristics remain open to find.

\section{Concluding remarks}\label{sec-4}

In this paper, we have shown how bilinear equations are converted into their nonlinear forms
by using the Bell polynomials.
The KdV-type and mKdV-type bilinear equations served as illustrative examples,
where the key roles are played by the formulae \eqref{2.29} and \eqref{2.30}.

Among the examples of the KdV type, a nonlinear form of the Hietartinta equation \eqref{Hie-eq}
has been found, which is given in \eqref{Hie-eq-n} (or \eqref{Hie-eq-n-uv}) and is not known before.
We also presented a nonlinear form of an integrable 4-dimensional equation in the KP family,
see \eqref{KP-4D-u} or \eqref{KP-4D-w}.
Note that the related bilinear equation \eqref{KP-4D} is interesting in the sense that
it allows elliptic $\tau$ functions but no need to introduce elliptic curve moduli parameters
in the bilinear equation
(see (4.36b) in \cite{LZ-JNS-2022}).
It can also be considered as a 4-dimensional extension of the bilinear BKP$(-1)$ equation \eqref{HS-a}.
In addition to the integrable equations, we also gave three non-integrable examples in Sec.\ref{sec-3-2},
although two of them used to be incorrectly studied as integrable equations.

Apart from the KdV-type equations, some integrable mKdV-type bilinear equations have been nonlinearized
in Sec.\ref{sec-3-3}. It is notable that the bilinear system \eqref{5.17}
is shown to connected with the breaking soliton mKdV equation \eqref{mkdv-b-3},
the study of which is much less compared with the breaking soliton KdV equation.
In addition, to our knowledge, \eqref{mkdv-b-5} and \eqref{mkdv-xyt} are new integrable systems.
The former maybe recognized as a 5th-order breaking soliton mKdV equation,
while for equation \eqref{mkdv-xyt}, it seems its integrability context remains unknown
except having 3-soliton solutions.

It is well known that integrable bilinear equations can be categorized according to different 
affine Lie algebras (e.g. \cite{JimboM-1983}) and toroidal Lie algebras 
(e.g. \cite{Billig-JA-1999,ISW-PLA-1999,KIT-AHP-2002,IT-IMRN-2001,YC-LMP-2024}),
but nonlinear forms of many bilinear equations are not yet known.
Since the Bell polynomials also can be used to construct
bilinear B\"{a}cklund transformations and then Lax pairs
for the KdV-type equation \cite{Lambert2001,LamS-2006},
one may keep using the Bell polynomials to study equations
\eqref{Hie-eq-n}, \eqref{KP-4D-w},  \eqref{mkdv-b-5} and \eqref{mkdv-xyt}
so as to achieve more insight of their integrability.
This will be considered in the future.
In addition, in this paper, we only investigated some bilinear KdV-type and mKdV-type equations.
It would be interesting to study nonlinearization of the integrable sG-type and
NLS-type bilinear equations presented in \cite{Hie-1987c,Hie-1988},
which involve complex conjugate operations
but the corresponding Bell polynomials approach is much less developed
compared with the wide application in the KdV-type equations.
Finally, note that the Bell polynomials are related to the Burgers hierarchy \cite{gilson-1996,lambert-1994}
and the discrete Burgers equation has been well understood in \cite{CZZ-SAPM-2021,Zhang-Burgers-2022}
recently. 
A related topic that might be interesting is whether there is a discrete version of the Bell polynomial approach
to study the discrete KdV type nonlinear equations
and also bilinear ones such as those having 3-soliton solutions found in \cite{HZ-JDEA-2013}.

\vskip 20pt
\subsection*{Acknowledgments}
This project is supported by the NSF of China (No.12271334).

\end{document}